\documentclass{ws-book9x6}
\usepackage{ws-book-thm}   
\usepackage{ws-book-har}   
\usepackage[pdfpagelabels=false,colorlinks=true,allcolors=black]{hyperref}  

\usepackage{psfrag}

\usepackage{tikz}
\usetikzlibrary{patterns,decorations,arrows,shapes.geometric,arrows,mindmap,backgrounds,positioning,positioning,fit,backgrounds,positioning,arrows,matrix,calc}
\usetikzlibrary{shapes,backgrounds,mindmap,shadows,shapes.geometric,topaths,snakes,arrows,pgfplots.groupplots,fadings,calc,decorations.markings,positioning,chains,fit}
\usetikzlibrary{arrows,shapes,positioning}
\usetikzlibrary{decorations.markings}
\tikzstyle arrowstyle=[scale=1]
\tikzstyle directed=[postaction={decorate,decoration={markings,
    mark=at position .65 with {\arrow[arrowstyle]{stealth}}}}]

\title{Sping Glass Theory and Far Beyond}      

\makeindex

\newcommand{\s}{\sigma}
\newcommand{\us}{\underline{\sigma}}
\newcommand{\da}{{\partial a}}

\newcommand{\di}{{\partial i}}
\newcommand{\dami}{{\partial a \setminus i}}
\newcommand{\dima}{{\partial i \setminus a}}

\newcommand{\e}{\varepsilon}
\newcommand{\tq}{\widetilde{q}}
\newcommand{\hmu}{\widehat{\mu}}
\newcommand{\heta}{\widehat{\eta}}
\newcommand{\hz}{\widehat{z}}
\newcommand{\hf}{\widehat{f}}
\newcommand{\hP}{\widehat{P}}
\newcommand{\hF}{\widehat{F}}

\newcommand{\eqd}{\overset{\rm d}{=}}
\newcommand{\g}{\gamma}
\newcommand{\Si}{\Sigma}
\newcommand{\dd}{{\rm d}}

\begin{document}

\setcounter{chapter}{17}

\chapter[The cavity method]{The cavity method: from exact solutions to algorithms}

\centerline{\large Alfredo Braunstein$^1$, Guilhem Semerjian$^2$}

\bigskip

\noindent
$^1$ Politecnico di Torino, Corso Duca degli Abruzzi, 24, I-10129, Torino, Italy,
Italian Institute for Genomic Medicine, IRCCS Candiolo, SP-142, I-10060, Candiolo (TO), Italy and INFN, Sezione di Torino, Italy

\medskip

\noindent
$^2$ Laboratoire de Physique de l'\'Ecole Normale Sup\'erieure, ENS, Universit\'e PSL, CNRS, Sorbonne Universit\'e, Universit\'e Paris Cit\'e, F-75005 Paris, France

\section{Introduction}

The quest of an analytic solution for the simplest mean-field spin-glass model (the Sherrington-Kirkpatrick (SK) one~\cite{SK75}) led Giorgio Parisi to the invention of the replica method~\cite{Pa80}. This method is able to describe and handle the complicated structure of the configuration space of the SK model, with a hierarchical division of the configurations into nested pure states, through the analytical parametrization of matrices of size $n \times n$, in the limit where $n\to 0$, which is, to say the least, a questionable mathematical construction (its predictions have been nevertheless confirmed rigorously later on~\cite{GuTo02,Ta06,Pa13}). In the physics literature an alternative method to solve the SK model was proposed in~\cite{MePaVi86}, and subsequently dubbed the cavity method. In a nutshell the idea is to consider the effect of the addition of one spin in a large SK model, or equivalently to create a ``cavity'' by isolating one spin and modeling the influence that the rest of the system has on it in a self-consistent way. The replica and the cavity methods yield the same predictions for the SK model, with complementary insights on its structure, the cavity method bypassing the ``analytic continuation'' from integer values of $n$ to $0$.

Even if the replica and cavity methods have had an impact inside physics, in particular in the context of structural glasses, they have also been very fruitful in fields which at first sight could seem unrelated, and in particular in computer science, information theory and discrete mathematics. Roughly speaking, the reason for their versatility lies in the rather universal character of the structure of the configuration space evoked above, that appears not only in the SK model but in many other problems with a non-physical origin, notably some random constraint satisfaction problems and error correcting codes. It turns out indeed that these problems can be viewed as mean-field spin-glasses, but slightly different from the SK one: the degrees of freedom in these problems interact strongly with a finite number of neighbors, whereas in the SK all degrees of freedom interact with each other weakly, in a ``fully-connected'' manner. The mean-field character of these sparse, or diluted, models arise from the choice of the neighbors, which is done uniformly at random, without the geometrical constraints of an Euclidean space. In physics terms such a network of interaction is called a Bethe lattice, in mathematics a random graph. This type of model appeared in the physics literature relatively shortly after the fully-connected ones~\cite{VB85}, but it became quickly clear that they were much more challenging to solve, some simplifications of the diverging connectivity (of a central limit theorem flavor) being absent in this case. A line of research extended the replica method to this sparse setting, see in particular~\cite{replica_diluted,BiMoWe} and references therein, at the price of a rather complicated order parameter. It turned out that the cavity method is a more convenient framework than the replica one for these problems, the complex configuration space encoded by the replica symmetry breaking being formulated in a more transparent manner through the cavity approach, as first discussed in~\cite{cavity}; in addition the formalism of the cavity method can be used to develop algorithms that provide informations on a single sample of mean-field spin-glasses, not only on average thermodynamic quantities.

The goal of this chapter is to review the main ideas that underlie the cavity method for models defined on random graphs, as well as present some of its outcomes, focusing on the random constraint satisfaction problems for which it provided both a better understanding of the phase transitions they undergo, and suggestions for the development of algorithms to solve them. It is organized as follows; section~\ref{sec_cavity} focuses on the analytic aspects of the method. It contains an introduction to models defined on random graphs (in Sec.~\ref{sec_rg}), then the equations of the cavity method at the so-called replica symmetric (RS) level and one step of replica symmetry breaking (1RSB) are presented in Sec.~\ref{sec_rs} and \ref{sec_rsb}, before reviewing in Sec.~\ref{sec_predictions} their outcomes concerning the phase diagram of random constraint satisfaction problems. Algorithmic consequences of this approach are detailed in Sec.~\ref{sec_algorithms}.

\section{The cavity method for sparse mean-field models}
\label{sec_cavity}

\subsection{Models on random graphs}
\label{sec_rg}

We shall consider systems made of $N$ elementary degrees of freedom (spins) $\s_i$, which take values in some finite alphabet $\chi$, and whose global configuration will be denoted $\us=(\s_1,\dots,\s_N) \in \chi^N$. They interact through an energy function (also called Hamiltonian, or cost function), that we decompose as
\begin{equation}
E(\us) = \sum_{a=1}^M \e_a(\us_{\da}) \ ,
\label{eq_energy}
\end{equation}
where the sum runs over the $M$ basic interactions terms $\e_a$. We denote $\partial a \subset \{1,\dots,N\}$ the set of variables involved in the $a$'th constraint, and for a subset $S$ of the variables $\us_S$
means $\{\s_i | i \in S \}$. In what follows we assume that all interactions involves a subset of $k$ variables, for a given $k \ge 2$. This framework encompasses usual Ising spin-glass models, with $\chi=\{-1,1\}$, $k=2$ and $\e_a(\us_{\da})=-J_a \s_{i_a} \s_{j_a}$, $J_a$ being the coupling constant between the spins $i_a$ and $j_a$. It also allows to deal with Potts spins when $\chi=\{1,\dots,q\}$ for a number $q\ge 2$ of spin states, also interpreted as colors; in this case a relevant energy function corresponds to pairwise interactions ($k=2$), with $\e_a(\us_{\da})=\delta_{\s_{i_a}, \s_{j_a}}$. This yields the Hamiltonian of the Potts antiferromagnetic model, corresponding in the perspective of computer science to the $q$-coloring problem, the cost function counting the number of monochromatic edges among the interacting ones. More generically a constraint satisfaction problem (CSP) corresponds to a cost function of the form (\ref{eq_energy}) with $\e_a$ taking values $0$ or $1$, and being interpreted as the indicator function of the event ``the $a$-th constraint is not satisfied by the configuration of the variables in $\us_{\da}$''. In particular the $k$-SAT and $k$-XORSAT problems can be described in this way with Ising spins and $k$-wise interactions. One calls solution of a CSP a configuration $\us$ satisfying simultaneously all the constraints, i.e. a zero-energy groundstate, and one says that the CSP is satisfiable if and only if it admits at least one solution.
 
The Gibbs-Boltzmann probability measure associated to this Hamiltonian for an inverse temperature $\beta$ reads
\begin{equation}
\mu(\us)= \frac{1}{Z} \prod_{a=1}^M w_a(\us_\da)  \ ,
\quad 
Z = \sum_{\us \in {\cal X}^N} \prod_{a=1}^M w_a(\us_\da)
\ , \quad
\Phi = \frac{1}{N} \ln Z \ .
\label{eq_mu_first}
\end{equation}
where the partition function $Z$ ensures the normalization of the probability 
law, and $w_a(\us_\da)=e^{-\beta \e_a(\us_\da)}$. We introduced the thermodynamic potential $\Phi$ which we shall call a free-entropy, as we did not include the constant $-1/\beta$ that would make it a free-energy. This choice allows to handle the uniform measure over the solutions of a CSP (assumed to be satisfiable), that corresponds to $w_a(\us_\da)=(1-\e_a(\us_\da))$, in which case $Z$ counts the number of solutions and $\Phi$ is the associated entropy rate. It amounts to set formally $\beta=\infty$ in the Gibbs-Boltzmann definition, in other words to work directly at zero temperature.

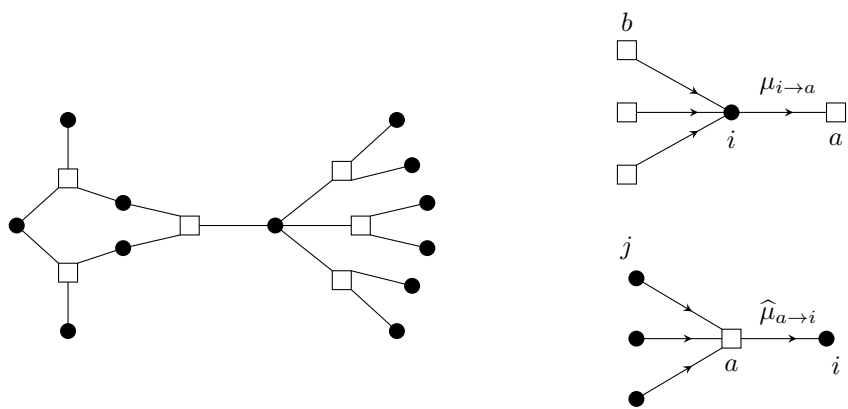
\begin{figure}
\begin{tikzpicture}
\fill[black] (0,0) circle (3pt);
\draw (1,-0.125) rectangle (1.25,0.125);
\draw (.75,0.6) rectangle (1,.85);
\draw (.75,-0.6) rectangle (1.,-.85);
\fill[black] (2,0.3) circle (3pt);
\fill[black] (2,-0.3) circle (3pt);
\fill[black] (1.8,0.8) circle (3pt);
\fill[black] (1.6,1.4) circle (3pt);
\fill[black] (1.8,-0.8) circle (3pt);
\fill[black] (1.6,-1.4) circle (3pt);
\draw (-1,-0.125) rectangle (-1.25,0.125);
\fill[black] (-2,0.3) circle (3pt);
\fill[black] (-2,-0.3) circle (3pt);
\draw (-1,0) -- (1,0);
\draw (0,0) -- (.75,.6);
\draw (0,0) -- (.75,-.6);
\draw (1.25,0.125) -- (2,0.3);
\draw (1.25,-0.125) -- (2,-0.3);
\draw (1,.85) -- (1.6,1.4);
\draw (1,.6) --  (1.8,0.8);
\draw (1,-.85) -- (1.6,-1.4);
\draw (1,-.6) --  (1.8,-0.8);
\draw (-1.25,0.125) -- (-2,0.3);
\draw (-1.25,-0.125) -- (-2,-0.3);
\fill[black] (-3.4,0.) circle (3pt);
\draw (-2.6,0.5) rectangle (-2.85,0.75);
\fill[black] (-2.725,1.4) circle (3pt);
\draw (-2.6,-0.5) rectangle (-2.85,-0.75);
\fill[black] (-2.725,-1.4) circle (3pt);
\draw (-2.725,1.4) -- (-2.725,.75);
\draw (-3.4,0.) -- (-2.85,0.5);
\draw (-2,0.3) -- (-2.6,0.5);
\draw (-2.725,-1.4) -- (-2.725,-.75);
\draw (-3.4,-0.) -- (-2.85,-0.5);
\draw (-2,-0.3) -- (-2.6,-0.5);
\begin{scope}[xshift=6cm,yshift=1.5cm]
\fill[black] (0,0) circle (3pt);
\draw (1.25,-0.125) rectangle (1.5,0.125);
\draw[directed] (0,0) -- (1.25,0);
\draw (-1.25,-0.125) rectangle (-1.5,0.125);
\draw (-1.25,0.7) rectangle (-1.5,0.95);
\draw (-1.25,-0.7) rectangle (-1.5,-0.95);
\draw[directed] (-1.25,0) -- (0,0);
\draw[directed] (-1.25,0.7) -- (0,0);
\draw[directed] (-1.25,-0.7) -- (0,0);
\draw (0.,-.35) node {$i$};
\draw (1.375,-.35) node {$a$};
\draw (.75,.35) node {$\mu_{i \to a}$};
\draw (-1.375,1.2) node {$b$};
\end{scope}
\begin{scope}[xshift=6cm,yshift=-1.5cm]
\draw (-.125,-0.125) rectangle (.125,0.125);
\fill[black] (1.25,0) circle (3pt);
\draw[directed] (0.125,0) -- (1.25,0);
\fill[black] (-1.25,0) circle (3pt);
\fill[black] (-1.25,0.8) circle (3pt);
\fill[black] (-1.25,-0.8) circle (3pt);
\draw[directed] (-1.25,0) -- (-0.125,0);
\draw[directed] (-1.25,0.8) -- (-0.125,0.125);
\draw[directed] (-1.25,-0.8) -- (-0.125,-0.125);
\draw (0.,-.35) node {$a$};
\draw (1.375,-.35) node {$i$};
\draw (.75,.35) node {$\widehat{\mu}_{a \to i}$};
\draw (-1.375,1.2) node {$j$};
\end{scope}
\end{tikzpicture}

\caption{
Left: an example of a factor graph. 
Right: illustration of Eqs.~(\ref{eq_BP_mu},\ref{eq_BP_hmu}).}
\label{fig_fg}
\end{figure}

A convenient representation of a probability measure $\mu$ of the form (\ref{eq_mu_first}) is provided by a factor graph~\cite{factorgraph}, see Fig.~\ref{fig_fg} for an example, which is a bipartite
graph where each of the $N$ variables $\s_i$ is represented by a circle
vertex, while the $M$ weight functions $w_a$ are associated to square vertices. An edge
is drawn between a variable $i$ and an interaction $a$ if and only if 
$w_a$ actually depends on $\s_i$, i.e. $i \in \da$. In a similar way
we shall denote $\partial i$ the set of interactions in which $\s_i$ appears,
i.e. the graphical neighborhood of $i$ in the factor graph, and call $|\di|$ the degree of the $i$-th variable. One has a
natural notion of graph distance between two variable nodes $i$ and $j$,
defined as the minimal number of interaction nodes on a path linking $i$ and
$j$.

Our interest lies in disordered systems, in which the probability measure $\mu$ is itself a random object. Suppose indeed that the weight functions $w_a$ are built by drawing, independently for each $a$, the $k$-uplet of variables $\da$ uniformly at random among the $\binom{N}{k}$ possible choices (and also the coupling constants defining the interaction if necessary). We will denote $\mathbb{E}[\bullet]$ the average with respect to this quenched randomness (let us emphasize that there are two distinct level of probabilities in these systems: the spins $\us$ are random variables with the probability law $\mu$, and $\mu$ is random because of the stochastic choices in the construction of the factor graph).  For $k=2$ the resulting factor graph is drawn from nothing but the celebrated Erd{\H{o}}s-R{\'e}nyi $G(N,M)$ random graph ensemble, the case $k>2$ corresponding to its natural hypergraph generalization. The large size (thermodynamic) limit we shall consider corresponds to $N,M\to \infty$, with $\alpha= M/N$ a fixed parameter. Let us recall some elementary properties of these random factor graphs in this limit:
\begin{itemize}
\item the probability that a randomly chosen variable $i$ has degree 
$|\di|=d$ is $q_d=e^{-\alpha k} (\alpha k)^d/d!$, the Poisson law of
mean $\alpha k$.
\item if one chooses randomly an interaction $a$, then a variable $i\in\da$,
the probability that $i$ appears in $d$ interactions \emph{besides} $a$, i.e. 
that $|\di \setminus a|=d$, is $\tq_d =e^{-\alpha k} (\alpha k)^d/d!$.
\item the random factor graphs are locally tree-like: choosing at random a 
vertex $i$, the subgraph made of all nodes at graph distance from $i$ smaller 
than some threshold $t$ is, with a probability going to 1 in the thermodynamic
limit with $t$ fixed, a tree.
\end{itemize}

More general ensembles of random factor graphs can be constructed, by fixing
a degree distribution $q_d$ and drawing at random from the set of all graphs
of size $N$ with $N q_0$ isolated vertices, $N q_1$ vertices of degree 1, and
so on and so forth. Then the two distributions $q_d$ and $\tq_d$ are
different in general, and related through 
$\tq_d = (d+1) q_{d+1}/\sum_{d'} d' q_{d'}$. An important example in this class
corresponds to random regular graphs, where $q_d$ is supported by a single 
integer.

\subsection{The replica symmetric (RS) cavity method}
\label{sec_rs}

The goal of the cavity method is to describe the properties of the random measure $\mu$ constructed above, for typical samples of the random graph ensemble. The free-entropy $\Phi$ is self-averaging in the thermodynamic limit, its typical value concentrates around its average, the quenched free-entropy $\phi$ defined as
\begin{equation}
\phi = \lim_{N \to \infty} \mathbb{E}[\Phi] = \lim_{N \to \infty}\frac{1}{N} \mathbb{E}[ \ln Z] \ .
\label{eq_def_phi}
\end{equation}
The computation of this quantity is thus the objective of the cavity method, along with a local description of the measure $\mu$, in terms of its marginal distributions on a finite number of spins.

The cavity method relies crucially on the local convergence of random factor
graph models to random trees explained at the end of Sec.~\ref{sec_rg}.
Let us assume momentarily that the factor graph representing 
the model under study is a finite tree. Then the problem of characterizing the 
measure~(\ref{eq_mu_first}) and computing the associated partition function $Z$
can be solved exactly in a simple, recursive way: one can break the tree into independent subtrees, solve the problems on these substructures, and combine them together to get the solution on the larger problem. This is nothing but a generalization of the transfer matrix method used in physics to solve unidimensional problems, a form of what is known as dynamic programming in computer science. More precisely, for each edge between a variable $i$ and an adjacent interaction $a$ one introduces two directed ``messages'', $\mu_{i \to a}$ and $\hmu_{a \to i}$, which are probability measures on the alphabet $\chi$, that would be the marginal probability of $\s_i$ if, respectively, the interaction $a$ were removed from the graph, or if all interactions around $i$ except $a$ were removed. A moment of thought reveals that these messages obey the following recursive (so-called Belief Propagation (BP)) equations (see the right part of Fig.~\ref{fig_fg} for an illustration),
\begin{align}
\mu_{i \to a}(\s_i) &=  \frac{1}{z_{i \to a}}  \prod_{b\in \dima} \hmu_{b \to i}(\s_i) \ , \label{eq_BP_mu} \\
\hmu_{a \to i}(\s_i) &= \frac{1}{\hz_{a \to i}} \sum_{\us_\dami} w_a(\us_\da)
\prod_{j \in \dami} \mu_{j \to a}(\s_j) \ , 
\label{eq_BP_hmu}
\end{align}
with $z_{i \to a}$ and $\hz_{a\to i}$ ensuring the normalization of the laws.
On a tree factor graph there exists a single solution of these equations, which
is easily determined starting from the leaves of the graph (for which the empty
product above is conventionally equal to 1) and sweeping towards the inside of
the graph. Once the messages have been determined all local
averages with respect to $\mu$ can be computed, as well as the partition 
function, in terms of the solutions of these BP equations. 
The Belief Propagation algorithm consists in looking for a fixed-point solution
of (\ref{eq_BP_mu},\ref{eq_BP_hmu}), iteratively, even if the factor graph is not a tree; in this case the formula giving $\Phi$ in terms of the messages is only an approximation, known as the Bethe formula for the free-entropy (see for instance~\cite{Yedidia2} for more details on the connections between the stationary points of the Bethe free-entropy and the solutions of the Belief Propagation equations). These equations were discovered independently in Statistical Physics as the Bethe-Peierls
approximation, in artificial intelligence as the Belief Propagation algorithm, and in Information Theory as the Sum-Product algorithm~\cite{MM09}.

Of course random graphs are only locally tree-like, they do possess loops,
even if their lengths typically diverge in the thermodynamic limit. The
cavity method amounts thus to a series of prescriptions to handle these long
loops and to describe the boundary condition they impose on the local tree
neighborhoods inside a large random graph. The simplest prescription, that
goes under the name of replica symmetric (RS) and that is valid for weakly interacting models (i.e. small $\alpha$ and/or large temperature), assumes some spatial correlation decay properties of the probability measure $\mu$. When one removes an interaction $a$ from a factor graph the variables around it becomes strictly independent if one starts from a tree, and asymptotically independent provided only long enough loops join them in absence of $a$, and provided the correlation decays fast enough along these loops. To
compute the average thermodynamic potential (\ref{eq_def_phi}) it is enough in
this case to study the statistics with respect to the quenched disorder
of the messages $\mu_{i \to a}$, $\hmu_{a \to i}$ on the edges of the random factor graph.
In other words the order parameter of the RS cavity method is the law of the random variables $\eta$, $\heta$, which are equal to the random messages one obtains by drawing at random a sample, solving the BP equations on it, choosing at random an edge $a-i$, and observing the value of $\mu_{i \to a}$ and $\hmu_{a \to i}$. With the assumption of independence underlying the RS cavity method the equations (\ref{eq_BP_mu},\ref{eq_BP_hmu}) translate into Recursive Distributional Equations (RDE) of the form:
\begin{equation}
\eta \eqd 
f(\heta_1,\dots,\heta_d) \ , \qquad \heta \eqd \hf(\eta_1,\dots,\eta_{k-1})
\ .
\label{eq_eta_RSclass}
\end{equation}
In this equation all the $\eta_i$'s and $\heta_i$'s are independent copies of the random variables $\eta$ and $\heta$, $\eqd$ denotes the equality in distribution between random variables, $d$ is drawn according to the law $\tq_d$, and the functions $f$ and $\hf$ are defined by the right hand sides of equations (\ref{eq_BP_mu},\ref{eq_BP_hmu}) (with possibly an additional random draw of the weight $w$).
The RS prediction for
$\phi$ can then be expressed as the average over random copies of $\eta$ and $\heta$ of the local free-entropy contributions obtained from the exact computation of the partition function of
a finite tree. Note that the equation (\ref{eq_eta_RSclass}), if it has in
general no analytic solution, lends itself to a very natural numerical
resolution where the law of $\eta$ is approximately represented as an empirical 
distribution over a set of representatives $\eta$ (a population 
representation)~\cite{abou1973,cavity}.

The exactness of the predictions of the RS cavity method has been proven rigorously for some models which are not too frustrated (e.g. ferromagnetic systems, or matching models), see for instance ~\cite{DM10,BoLe10,BoLeSa_matchings}. But in general the correlation decay assumption fails, in this case one has to turn to a more sophisticated version of the cavity method, which will be introduced in the next section.


\subsection{Handling the replica symmetry breaking (RSB) with the cavity method}
\label{sec_rsb}

As a matter of fact for low enough temperature, and high enough density of interactions $\alpha$, the configuration space of frustrated random models gets fractured in a large number of pure states (or clusters), and the correlation decay hypothesis only holds for the Gibbs measure restricted to one pure state, not for the 
complete Gibbs measure. In the replica method this phenomenon shows up as a breaking of the equivalence between different replicas, we will now explain how the cavity method is able to handle this structure of the configuration space.
It amounts to make further self-consistent hypotheses on the correlated boundary 
conditions this induces on the tree-like portions of the factor graph.
Inside each pure state the RS computation is assumed to hold true, and the RSB computation
is then a study of the statistics of the pure states. Let us explain
how this is done in practice at the first level of RSB (1RSB cavity method).
The partition function is written as a sum over the pure states $\gamma$,
that form a partition of the configuration space,
$Z=\sum_\g Z_\g$, where $Z_\g$ is the partition function restricted to
the pure state $\g$. It can be written in the thermodynamic limit as
$Z_\g = e^{N f_\g}$, with $f_\g$ the internal free-entropy density of a given pure state. One further assumes 
that the number of pure states with 
a given value of $f$ is, at the leading 
exponential order, $e^{N\Si(f)}$, with the so-called configuration entropy, or complexity, $\Si$ 
a concave function of $f$,
positive on the interval $[f_{\rm min},f_{\rm max}]$. In order to compute
$\Si$ one introduces a parameter $m$ (called Parisi breaking parameter)
conjugated to the internal thermodynamic potential, 
and the generating function of the
$Z_\g$ as ${\cal Z}(m) = \sum_\g Z_\g^m$. In the thermodynamic limit its dominant 
behavior is captured by the 1RSB potential $\phi_{\rm 1RSB}(m)$,
\begin{equation}
\phi_{\rm 1RSB}(m)= \lim_{N \to \infty} \frac{1}{N}\log {\cal Z}(m) = 
\sup_f \left[\Si(f)+ m f \right] \ ,
\label{eq_Phi_1RSBclass}
\end{equation}
where the last expression is obtained by a saddle-point evaluation of the
sum over $\g$. The complexity function is then accessible via the inverse
Legendre transform of $\phi_{\rm 1RSB}(m)$~\cite{Mo95}, or in a parametric form
\begin{equation}
f(m) = \phi_{\rm 1RSB}'(m)  \ , \qquad
\Si(f(m)) = \phi_{\rm 1RSB}(m) - m \phi_{\rm 1RSB}'(m) \ , 
\end{equation}
where $f(m)$ denotes the point where the supremum is reached in 
Eq.~(\ref{eq_Phi_1RSBclass}). One has $\Si'(f(m))=-m$, i.e. the
introduction of the parameter $m$ allows to explore the complexity
curve by tuning the tangent 
slope of the selected point.

The actual computation of $\phi_{\rm 1RSB}(m)$
is done as follows~\cite{cavity}. One introduces on each
edge of the factor graph two distributions $P_{i \to a}$ and $\hP_{i \to a}$ of messages,
which are the probability over the different pure states $\gamma$, weighted
proportionally to $Z_\gamma^m$, to observe a given value of $\mu_{i \to a}^\gamma$ and $\hmu_{a \to i}^\gamma$ respectively,
where $\mu_{i\to a}^\gamma$ and $\hmu_{a \to i}^\gamma$ are the messages that appear in Eq.~(\ref{eq_BP_mu},\ref{eq_BP_hmu}),
for the measure restricted to the pure state $\gamma$.
Because $P_{i \to a}$ and $\hP_{a \to i}$ are themselves random objects with respect to the choices in the generation of the instance of the factor graph,
the order parameter of the 1RSB cavity method becomes the distributions of $P_{i \to a}$ and $\hP_{a \to i}$ with respect to the disorder. The latter is solution of a self-consistent
functional equation written as
\begin{equation}
P \eqd F(\hP_1,\dots,\hP_d) \ , \qquad 
\hP \eqd \hF(P_1,\dots,P_{k-1}) \ ,
\label{eq_P_1RSBclass}
\end{equation}
that parallels the equation (\ref{eq_eta_RSclass}) of the RS cavity method, with again independent copies of the distributions $P_i$ and $\hP_i$.
The right hand sides of these distributional equalities stand for:
\begin{align}
P(\eta) &= \frac{1}{Z} \int \prod_{i=1}^d \dd \hP_i(\heta_i) \ \delta(\eta-f(\{ \heta_i \})) \
z(\{ \heta_i \})^m \ , \label{eq_1RSB_P} \\
\hP(\eta) &= \frac{1}{\widehat{Z}} \int \prod_{i=1}^{k-1} \dd P_i(\eta_i) \ \delta(\heta-\hf(\{ \eta_i \})) \
\hz(\{ \eta_i \})^m \ , 
\label{eq_1RSB_hP}
\end{align}
with the functions $f$ and $\hf$ corresponding to the recursion functions at the RS level, see Eq.~(\ref{eq_BP_mu},\ref{eq_BP_hmu}), and $z$ and $\hz$ the associated normalization factors.
From the solution of this equation 
(that again can be found numerically with the population dynamics 
method~\cite{cavity}) one
computes the 1RSB potential $\phi_{\rm 1RSB}(m)$ via an expression similar to
the one giving the expression of $\phi$ at the RS level, with now averages
over random distributions $P$ and $\hP$.

There are different justifications for the appearance of the 
``reweighting factors'' $z^m$ and $\hz^m$ in 
Eqs.~(\ref{eq_1RSB_P},\ref{eq_1RSB_hP}). The argument in~\cite{cavity} is based on the
exponential distribution of the free-entropies $N f_\gamma$ of the pure states
with respect to some reference value, and on consistency requirements
on the evolution of the pure states when the cavity factor graph is modified. 
One can also study the statistics
of the many fixed point solutions of the BP equations (\ref{eq_BP_mu},\ref{eq_BP_hmu}) and
devise a dual factor graph for the counting of these fixed 
points~\cite{MM09}, the reweighting factor allowing to select the fixed
points associated to some internal free-entropy. Another interpretation
was proposed in~\cite{KrMoRiSeZd}, associating the pure states of a large
but finite factor graph model to boundary conditions on trees. This interpretation is particularly relevant in the case $m=1$, for which these boundary conditions are actually drawn from the Gibbs measure itself, and reveals a deep connection between the 1RSB cavity method and the reconstruction on tree problem, as first unveiled in~\cite{MM06}, and with the point-to-set correlations of the Gibbs measure~\cite{MoSe2}.

This construction can be generalized to higher levels of replica symmetry breaking~\cite{Pa80}, with a hierarchical partition of the configuration space into nested pure spaces; the resulting equations for models on sparse random graphs involve a recursive tower of probability distributions over probability distributions, whose numerical resolution becomes extremely challenging beyond 1RSB.

\subsection{Some analytic outcomes of the cavity method}
\label{sec_predictions}

As presented above the cavity method is quite versatile, in the sense that it can address a variety of models defined on random graphs, and it has indeed been applied to several different problems. As an illustration of some of its outcomes we shall now present some results it has provided on the phase diagram of random constraint satisfaction problems (see also chapter 31), and sketch the connections between this qualitative understanding and the quantitative formalism we have introduced before.

In the case of a constraint satisfaction problem the cost function defined in Eq.~(\ref{eq_energy}) is made of a sum of indicator functions of events that the $a$-th constraint is unsatisfied, for instance the number of monochromatic edges in the $q$-coloring problem. The natural questions in this context are: does an instance of the problem admit at least one solution? if yes, how are the solutions organized in the configuration space? It turns out that the answers to these questions have drastically different answers depending on the value of the density of constraints $\alpha$, in other words there exist, in the thermodynamic limit, sharp phase transitions for some threshold values of this parameter.

\begin{figure}
\centerline{\includegraphics[width=.9\textwidth]{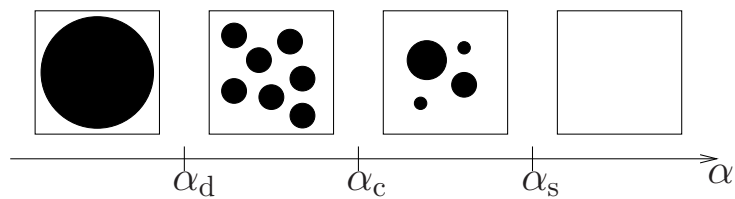}}
\caption{Schematic representation of the phase transitions in a random CSP
ensemble.}
\label{fig_clusters}
\end{figure}

The main transitions that occur for generic ensembles of random CSPs are represented in a schematic way on Fig.~\ref{fig_clusters}. The squares represent the full configuration space, for four different values of $\alpha$ (obviously the representation of this $N$-dimensional hypercube on a two-dimensional
drawing is only a cartoon), while the black area stands for the solutions. For $\alpha > \alpha_{\rm s}$, the satisfiability transition, the square is empty, which translates the absence of solution in typical instances for these density of constraints. The satisfiable regime $\alpha < \alpha_{\rm s}$ is further divided in three regions, separated by structural phase transitions at which the organization of the set of solutions changes qualitatively. For $\alpha<\alpha_{\rm d}$, the so-called clustering, or dynamic transition, all solutions
are somehow close to each other, while in the rest of the satisfiable regime they are broken in clusters of nearby solutions, each cluster being separated from the other ones. The number and size of the relevant clusters further change at the condensation threshold $\alpha_{\rm c}$: for $\alpha_{\rm d}<\alpha < \alpha_{\rm c}$ most solutions are contained in an exponential number of clusters which have all roughly the same size, while in the regime $\alpha_{\rm c}<\alpha < \alpha_{\rm s}$ most solutions are found in a sub-exponential number of clusters with strongly fluctuating sizes.

These qualitative predictions, along with quantitative numerical values for some definite random CSPs families, have been obtained by the analysis of the solutions of the 1RSB cavity equations, according to the following criteria~\cite{MeZe,KrMoRiSeZd}:
\begin{itemize}
\item $\alpha_{\rm d}$ is the smallest value of $\alpha$ such that the 1RSB equations at $m=1$ admit a non-trivial solution.

\item in the regime $[\alpha_{\rm d},\alpha_{\rm c}]$ the configurational entropy, or complexity, associated to the $m=1$ solution, is positive, whereas it becomes negative for $\alpha > \alpha_{\rm c}$.

\item the satisfiability transition is marked by the vanishing of the complexity computed at $m=0$, in the so-called energetic version of the 1RSB cavity method ~\cite{cavity_T0}, that counts all clusters irrespectively of their sizes.

\end{itemize}

%

\section{Some algorithmic outcomes of the cavity method}
\label{sec_algorithms}

\subsection{Algorithmic applications of the cavity method}

As mentioned above the equations \eqref{eq_BP_mu}-\eqref{eq_BP_hmu} can be used on a single instance to compute (approximately) several properties of the  distribution \eqref{eq_mu_first}, including single-site marginals,
joint marginals of variables in a common factor, the free energy and
Shannon's entropy. This approach has been applied to Bayesian networks, in the decoding phase of
communication codes (syndrome-based decoding, Turbo Codes \cite{benedetto_soft-output_1996})
and in stereo image reconstruction. More recently, it has found applications
in a large variety of fields that we shall now review. 

\paragraph{BP applications in notable models}

In \cite{kabashima_cdma_2003}, a Belief Propagation algorithm for
CDMA decoding has been presented. Interestingly, it shows how BP can
be efficiently applied to dense models (i.e. in which constraints
involve an extensive number of variables) through an application of
the Central Limit Theorem (the basis of a BP derivative called AMP, see Chapter 19), and it is also shown that solutions are
also fixed points of the famous Thouless-Anderson-Palmer (TAP) equations~\cite{TAP77}
while showing superior iterative convergence properties. A similar
approach has been employed in~\cite{braunstein_learning_2006} for
the binary discrete perceptron learning problem.

In \cite{frey_mixture_2005}, the Affinity Propagation (AP) algorithm
was presented. AP is a BP algorithm for variables with an extensive
number of states. The AP algorithm solves approximately a clustering
problem which is similar in spirit to $K$-means, but with the important
difference of only relying on a distance matrix instead of the original,
possibly high-dimensional, data representation. Auxiliary variables
with a large number of states can be employed to locally enforce global
constraints such as connectivity, by representing in the variables
state the discrete time of an underlying dynamics. BP has been applied
to the resulting extended model \cite{bayati_statistical_2008}.

The dynamic cavity method \cite{neri_cavity_2009} is an application of BP to
study a certain class of out-of equilibrium dynamical models. The
method can be understood as an application of BP to an auxiliary model
in which a variable consists in a couple of time-dependent quantities: one is a single spin trajectory,
the other a local field. Subsequent works showed that a slightly
simpler but equivalent representation can be obtained with a pair
of spin trajectories. On certain models such as discrete, microscopically
irreversible ones (i.e. ones in which a variable can never go back
to a visited state, including the Bootstrap percolation model \cite{altarelli_optimizing_2013},
SI or SIR epidemic models \cite{altarelli_bayesian_2014}), single
trajectories can be efficiently represented by the transition times.
In other cases, some approximations must be employed \cite{aurell_dynamic_2012}. A somehow related variant of the cavity method deals with quantum models, the basic degrees of freedom becoming imaginary-time spin trajectories~\cite{our_review_qaa}.

\paragraph{Exactness of BP on single instances}

Some rigorous results have been proven regarding the exactness of BP algorithms. For certain models and sufficiently large temperature,
the BP update equation becomes a contractive mapping, guaranteeing
the existence and uniqueness of its fixed point and the convergence towards
it under iterations thanks to the Banach theorem. Moreover, this condition
guarantees exactness in the thermodynamical limit on graphs with large
girth \cite{bayati_rigorous_2006}.

On the other side of the spectrum, some exactness results exist in
the small temperature limit as well. Equations to analize models explicitely
at zero temperature can be devised by taking the $T\to0$ limit of the BP equations under
an an opportune change of variables, resulting in equations for energy-shifts
instead of probabilities. These had been known in coding theory as Max-Sum algorithms. Existing proofs of exactness (on
some models) rely on a local optimality condition for BP fixed points.
\cite{bayati_maximum_2005,weiss_optimality_2001,gamarnik_belief_2012}.

Gaussian BP (GaBP) \cite{weiss_correctness_2001} is an application
of BP for a continuous model with positive definite quadratic potential,
i.e. a Multivariate Gaussian. It is shown under certain conditions on
the precision matrix that the GaBP equations converge and give the correct
estimation of the means (but wrong estimation of the variances in
general), effectively solving a linear system iteratively, with convergence
properties that make the method competitive. Note that due to the
fact that the mode is equal to the mean in a Gaussian distribution,
this result can be again thought of as the exactness of the computation of
the maximum.

\paragraph{Survey Propagations and the RSB Phase}

Survey propagation (SP) is the algorithmic counterpart of the 1RSB cavity
method. It has seen its first applications to study the $k$-SAT \cite{mezard_random_2002,braunstein_survey_2005}
and $q-$coloring \cite{col1} problems in the replica symmetry broken phase. SP can
be thought as BP for the combinatorial problem of solutions of a lower
order message passing system (typically Max-Sum or some coarsened
version of it). Such a hierarchical approach can also be employed to analyze problems that possess explicitely such a nested structure, such as the ones coming from
(stochastic) control problems (e.g. the Stochastic Matching problem
\cite{altarelli_stochastic_2011}).

It should also be noted that BP can be used in the RSB phase of constraint
satisfaction problems. In \cite{braunstein_learning_2006} BP has
been applied successfully to the perceptron learning problem with
binary synapses, even in the regime in which it shows a RSB phase.
The solution to this conundrum has been clarified in \cite{baldassi_unreasonable_2016},
where it was shown that BP describes an exponentially small portion
of the solution space that is still exponentally large and has a non-clustered
geometry akin to the dominant region of the solution space in the RS phase.

\paragraph{Decimation and reinforcement.}

An algorithm estimating marginal distributions such as BP can be employed
for sampling, and in particular to find solutions to a constraint
satisfaction problem. The main idea is ancestral sampling, i.e. given
an arbitrary permutation of variable indices $\pi$, one can estimate the marginal distribution
$p\left(x_{\pi_{1}}\right)$ and sample $x_{\pi_{1}}^{*}$ from it,
then restrict the solution space to solutions with $x_{\pi_{1}}=x_{\pi_{1}}^{*}$
and reiterate, effectively sampling $x_{\pi_{i}}^{*}\sim p\left(x_{\pi_{i}}|x_{\pi_{1}}^{*},\dots,x_{\pi_{i-1}}^{*}\right)$
for $i=1,\dots,n$. As $p\left(\underline{x}\right)=\prod_{i=1}^{n}p\left(x_{\pi_{i}}|x_{\pi_{1}},\dots,x_{\pi_{i-1}}\right)$,
this solution provides a fair sample $\underline{x}^{*}$ if 
the estimation of the marginals is exact. The analysis of ancestral sampling with BP has been
performed in \cite{Allerton,RiSe09,Coja11,Coja12}. When one is merely
interested in finding \textit{any} solution to a contraint satisfaction problem, and
remembering that marginal estimations are only approximate, it is
convenient to iteratively fix the variable that reduces the solution
space the \emph{less, }which corresponds to fixing the most polarized
variable in the direction of the largest probability of its marginal.
This process is called \emph{decimation}. In practice, decimation corresponds to iteratively selecting  the variable with the largest local field and applying an infinite external field to it  with the same sign (and then making the equations converge again and reiterating). A soft version of decimation, called reinforcement, can also be conceived, in which a field  is applied iteratively to all variables with the same sign of their local field and an intensity that is either a constant \cite{chavas_survey-propagation_2005} or proportional to its magnitude \cite{braunstein_learning_2006, bayati_statistical_2008}. This dynamics slowly drives the system to one with sufficiently large external fields that becomes trivially polarized on one solution. As an additional twist, a backtracking procedure can be implemented on top of decimation, in which variables are occasional freed from their external field when  that choice enlarges the solution space sufficiently. This has been implemented for SP, with excellent results \cite{marino_backtracking_2016}. 



\section{Conclusions}

The Cavity method is a powerful and versatile approach to the description of disordered systems, that has been shown so far to provide the exact asymptotic solution for many models. For given (finite) system instances, its algorithmic counterpart has many practical applications, ranging from a statistical description of the Boltzman-Gibbs distribution to the individuation of single solutions of a CSP. Moreover, at variance with more traditional methods for inference such as MCMC sampling, it can provide an analytical description, given implicitly by the solution(s) of the cavity equations. This fact enables many possibilities, such as its recursive application (SP), and a functional expression of statistical features as a function of the disorder parameters (see for instance chapter 21 for a discussion of inverse problems).




\bibliographystyle{ws-book-har}
\bibliography{biblio}

\end{document}